\documentclass{PoS}

\usepackage{amsmath}
\usepackage{amsfonts}
\usepackage{amssymb}
\usepackage{caption}
\usepackage{float}
\usepackage{graphicx}
\usepackage{slashed}
\usepackage{xfrac}

\newcommand{\p}{\vec{p}}
\newcommand{\pp}{\vec{p}\:\!'}
\newcommand{\q}{\vec{q}}
\newcommand{\x}{\vec{x}}
\newcommand{\cG}{G}

\newcommand{\chib}{\overline{\chi}}
\newcommand{\phib}{\overline{\phi}}

\title{Probing the proton and its excitations in full QCD}

\ShortTitle{Probing the proton and its excitations in full QCD}

\author{\speaker{Benjamin Owen}%
\thanks{We thank PACS-CS Collaboration for making their $2+1$ flavor
configurations available and acknowledge the important ongoing support
of the ILDG.  This research was undertaken with the assistance of
resources at the NCI National Facility in Canberra, Australia, and the
iVEC facilities at Murdoch University (iVEC@Murdoch) and the
University of Western Australia (iVEC@UWA). These resources were
provided through the National Computational Merit Allocation Scheme
and the University of Adelaide Partner Share which are supported by
the Australian Government.  We also acknowledge eResearch SA for their
support of our supercomputers.  This research is supported by the
Australian Research Council.}%
\:\,$^a$,\, Waseem Kamleh$^a$,\,  Derek Leinweber$^a$,\, Selim Mahbub$^{a,b}$ and Benjamin Menadue$^{a,c}$ \\ \\
\llap{$^a$}Special Research Centre for the Subatomic Structure of Matter (CSSM), \\
		School of Chemistry and Physics, University of Adelaide, SA 5005, Australia\\ \\
\llap{$^b$}CSIRO Computational Informatics, College Road, Sandy Bay, TAS 7005, Australia\\ \\
\llap{$^c$}National Computational Infrastructure, \\
		Australian National University, Canberra, ACT 0200, Australia \\ \\

        E-mail: \email{benjamin.owen@adelaide.edu.au}}

\abstract{ We present a first look at the application of variational
  techniques for the extraction of the electromagnetic properties of
  an excited nucleon system. In particular, we include preliminary
  results for charge radii and magnetic moments of the proton, its
  first even-parity excitation and the $\Delta^{+}$. }

\FullConference{31st International Symposium on Lattice Field Theory LATTICE 2013\\
                 July 29 -- August 3, 2013\\
                 Mainz, Germany}

\begin{document}

\section{Introduction}


In our recent work \cite{Owen:2012:PLB} we performed a calculation of
$g_A$ using the variational method.  We found that this
provides an automatic approach for suppressing excited state effects, 
eliminating the need for fine-tuning.  

Isolation of the eigenstates results in rapid ground state dominance.
This allows an earlier current insertion in three-point functions.
Fit windows commence earlier and are wider, suppressing excited-state
systematic error and reducing statistical uncertainties.  Similar
findings were presented for the rho meson in
Ref.~\cite{Owen:2012:POS}.

The use of the variational approach has the added benefit of
providing access to excited state matrix elements.  This is the
motivation of a growing number of studies that are moving beyond the
lowest energy eigenstates to better understand the dynamics of QCD.
Here we apply the correlation matrix to calculate the electromagnetic
form factors of the ground-state proton, its first even-parity
excitation and the $\Delta^+$.  From these we extract the charge
radius and magnetic moments for these states.  To the best of our
knowledge, this is the first determination of the electromagnetic
form factors of the first even-parity excitation of the nucleon.

\section{Correlation matrix methods for matrix elements}

The goal of this approach is to produce a set of operators
$\phi^{\alpha}$ that satisfy
\begin{equation}
\label{diag}
\langle \Omega | \phi^{\alpha} | \beta, p, s \rangle = \delta^{\alpha
  \beta} \, .
\end{equation}
The way in which this is achieved is to take an existing basis of operators $\lbrace \chi_i \rbrace$ and construct the operators $\phi^{\alpha}$ as a linear superposition
\[
\phi^{\alpha}(x) = \sum_{i} v^{\alpha}_{i}\,  \chi_{i}(x) \: , \hspace{20pt} \phib\,{}^{\alpha}(x) = \sum_{j} \chib_{j}(x)\, u^{\alpha}_{j} \, .
\]
Beginning with the matrix of cross correlators
\[
\cG_{ij}(\p, t) = \sum_{\x} e^{-i \p \cdot \x} \langle \Omega | \chi_i(x) \chib_j(0) | \Omega \rangle \, ,
\]
and noting $\cG_{ij}(\p, t)\, u^{\alpha}_{j}$ provides a recurrence
relation with time dependence $e^{-E_{\alpha} t}$, one can show, as demonstrated in \cite{Leinweber:2004}, that the necessary vectors $v^{\alpha}_{i}$ and $u^{\alpha}_{j}$ are the eigenvectors of the generalised eigenvalue equations
\begin{subequations}
	\begin{align}
	v^{\alpha}_{i} \: \cG_{ij}(\p, t_0 + \Delta t) \hspace{13pt} &
        = e^{-E_{\alpha} \Delta t} \: v^{\alpha}_{i} \: \cG_{ij}(\p,
        t_0) \, , \\
	\hspace{13pt} \cG_{ij}(\p, t_0 + \Delta t) \: u^{\alpha}_{j} & = e^{-E_{\alpha} \Delta t} \: \hspace{13pt} \cG_{ij}(\p, t_0) \: u^{\alpha}_{j} \, .
	\end{align}
\end{subequations}
It is worth noting that these equations are evaluated for a given
3-momentum $\p$ and so the corresponding operators satisfy
Eq.~\eqref{diag} for this momentum only. One can project out the
correlator for the state $\alpha$ by
\[
\cG^{\alpha}(\p, t) \equiv v^{\alpha}_{i}(\p) \: \cG_{ij}(\p, t) \: u^{\alpha}_{j}(\p) \, ,
\]
from which the desired quantities are extracted in the standard way.
To access the corresponding three-point correlator, it is a simple
matter of applying the relevant eigenvectors to the corresponding
three-point function, where care is taken to ensure that the
projection is done with the correct momenta for source and sink: 
\[
	\cG^{\alpha}(\pp, \p, t_2, t_1) \equiv v^{\alpha}_{i}(\pp) \: \cG_{ij}(\pp, \p, t_2, t_1) \: u^{\alpha}_{j}(\p) \, .
\]
From the projected two and three-point functions one then constructs a
suitable ratio in the standard way to isolate the desired matrix
element. Here we choose to use the ratio as defined in
\cite{Hedditch:2007}
\[
	R^{\alpha}(\pp, \p; \Gamma) = \sqrt{ \frac{\cG^{\alpha}(\pp, \p, t_2, t_1; \Gamma, \Gamma_4) \, \cG^{\alpha}(\p, \pp, t_2, t_1; \Gamma, \Gamma_4)}{\cG^{\alpha}(\pp, t_2; \Gamma_4) \, \cG^{\alpha}(\p, t_2; \Gamma_4)} } \, .
\]

To isolate the electric and magnetic form factors, we follow the
approach outlined in \cite{Leinweber:1990} for the proton, and
\cite{Leinweber:1992} for the $\Delta^+$ baryon.  In the following,
we have projected out the relevant correlators and constructed the
relevant ratios for the state $| \alpha \rangle$ and so will drop the
state label ($R \equiv R^{\alpha}$).  We also define the reduced
ratio $\overline{R}$ as
\[
\overline{R}^{\mu}(\pp,\p; \Gamma) = \left[ \frac{2 E_p}{E_p + M} \right]^{\sfrac{1}{2}} \left[ \frac{2 E_{p'}}{E_{p'} + M} \right]^{\sfrac{1}{2}} R^{\mu}(\pp,\p; \Gamma) \, .
\]

\section{Baryon Form Factors}


The electromagnetic interaction with a spin-$\sfrac{1}{2}$ system is
described by two independent form factors.
As we are interested in the static properties of these hadrons, we
shall consider the Sachs form factors $G_E$ (charge) and $G_M$
(magnetic). 
We choose the incoming state to be at rest and so as outlined in
Refs.~\cite{Leinweber:1990,Boinepalli:2006xd} we can extract $G_E$ and
$G_M$ through the following ratio terms:
\begin{equation}
G_E(q^2) = \overline{R}^{4}(\q, 0; \Gamma_4) \, , \hspace{20pt}
G_M(q^2) = \frac{(E + M)}{|\q_1|} \overline{R}^{3}(\q, 0; \Gamma_2) \, .
\end{equation}


The electromagnetic interaction with a spin-$\sfrac{3}{2}$ system is
described by four independent form factors.
Again we consider the Sachs form factors which are $G_{E0}$
(charge), $G_{M1}$ (magnetic-dipole), $G_{E2}$(electric-quadrupole)
and $G_{M3}$ (magnetic-octupole).
In this work will evaluate the charge radii and magnetic moments only.
Again, we choose the incoming state to be at rest and so as outlined in \cite{Leinweber:1992} we can extract $G_{E0}$ and $G_{M1}$ through the following ratio terms:
\begin{align*}
G_{E0} &= \frac{1}{3} \left( \overline{R}_{1}{}^{4}{}_{1}(\q, 0;
\Gamma_4) + \overline{R}_{2}{}^{4}{}_{2}(\q, 0; \Gamma_4) +
\overline{R}_{3}{}^{4}{}_{3}(\q, 0; \Gamma_4)\right) \, , \\
G_{M1} &= - \frac{3}{5} \frac{E+M}{|\q_1|}\left( \overline{R}_{1}{}^{3}{}_{1}(\q, 0; \Gamma_2) + \overline{R}_{2}{}^{3}{}_{2}(\q, 0; \Gamma_2) + \overline{R}_{3}{}^{3}{}_{3}(\q, 0; \Gamma_2) \right) \, .
\end{align*}

\subsection{Charge Radii and Magnetic Moments}

To extract the squared charge radius, we assume a dipole
Ansatz. 
For the magnetic moments, we assume that the charge and magnetic form
factors display common scaling at low $Q^2$ such that $G_M(0) =
G_M(Q^2)/G_E(Q^2)$ for a unit charge baryon and  the magnetic moment is
\begin{equation}
\label{mm}
\mu_B = G_M(0)\, \frac{e}{2M_B} = G_M(0)\, \frac{M_N}{M_B} \,
\frac{e}{2M_N} = G_M(0)\, \frac{M_N}{M_B} \, \mu_N \, ,
\end{equation}
where we choose to express the moments in nuclear magnetons, $\mu_N$.

\section{Calculation Details}

The key to success with the variational approach is to utilise a basis
of operators which provide a large span within excited state
spectrum. As there are a limited number of local operators for a given
$J^{PC}$, a great deal of work has been made by various groups in
increasing the available operators.  Here we choose to use
gauge-invariant Gaussian smearing of the fermion source and sinks as a
method of extending our operator basis \cite{Mahbub:2009,Mahbub:2010}.
We choose to use 16, 35, 100 and 200 sweeps of smearing in the spatial
dimensions with smearing fraction $\alpha = 0.7$ \cite{Mahbub:2009}
allowing for the construction of a 4 $\times$ 4 correlation matrix.
The proton and $\Delta^{+}$ are accessed with the standard interpolators
\[
\chi(x) = \epsilon^{a b c} \left( u^{T a}(x)\, C \gamma_5\, d^b(x) \right) u^c(x) \, ,
\hspace{20pt}
\mbox{and}
\hspace{20pt}
	\chi_{i}(x) = \epsilon^{a b c} \left( u^{T a}(x)\, C \gamma_{i}\, d^b(x) \right) u^c(x) \, ,
\]
respectively.
For the variational analysis we use $t_0 = 18$ and $\Delta t = 2$
relative to the source at $t=16$.  

The ensembles used in this calculation are the PACS-CS 2 + 1 flavour
dynamical-QCD gauge-field configurations \cite{Aoki:2008} made
available through the ILDG \cite{Beckett:2011}.  These configurations
are generated using an 
${\cal O}(a)$-improved Wilson-Clover fermion action and Iwasaki
gauge-action, with $\beta = 1.90$ resulting in a lattice spacing $a =
0.0907 \mathrm{fm}$. The lattices have dimension $32^3$ $\times$ $64$
giving rise to a spatial box of length $L = 2.9$ fm.  We consider four
masses for the ground state proton and $\Delta^+$ and the heaviest
two for the proton excitation.  For each mass, we have 350
configurations.

The quark source is at $t=16$ relative to a fixed boundary condition
defining $t=1$.  For the SST inversion we choose to use the fixed
current method with a conversed-vector current inserted at $t_S =
21$. For our error analysis we use a second-order single-elimination
jackknife method.  The $\chi^2$ per degree of freedom,
$\chi^2_{\textrm{dof}}$, is obtained via a covariance matrix
analysis.

\section{Results}

\begin{figure}[t]
\centering
\hspace*{-30pt}
\includegraphics[width=0.75\textwidth]{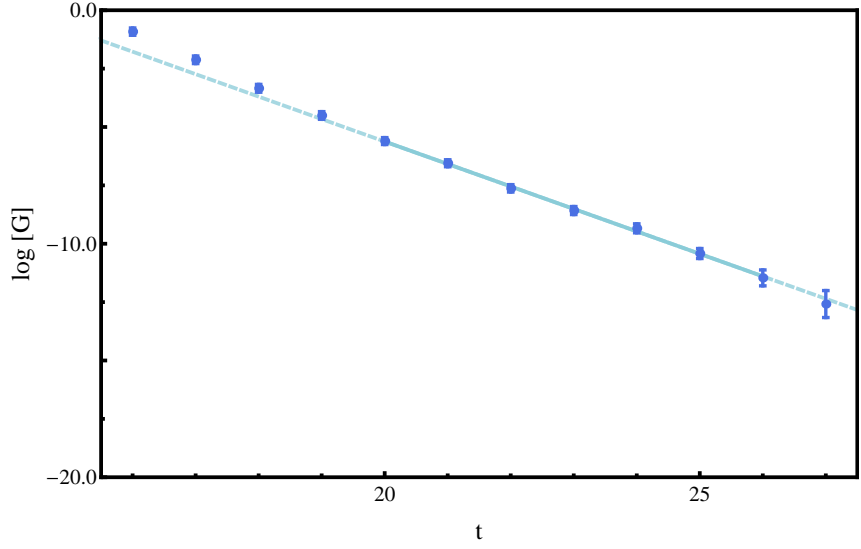}
\caption{$\mathrm{log}(G)$ for the projected correlator for the first
  even-parity excitation of the nucleon.  The solid line is the best fit
  line highlighting the linear behaviour of the correlator commencing
  at $t=20$ prior to the current insertion.  The correlator is
  consistent with a single isolated excited state over the region
  where the form factors will be analysed.}
\end{figure}

To ensure that we are in fact probing a nucleon excitation throughout
the Euclidean time range required, we consider the behaviour of the
projected two-point correlator.  In Fig.~1 we display
$\mathrm{log}(G)$ for the first excited state of the proton from the
variational analysis.  The linear behaviour observed between times
$t=20-26$ indicates that we have successfully isolated a single
eigenstate by $t=20$ (prior to the current insertion at $t_S = 21$)
and that this eigenstate dominates the correlator over the region
considered in calculating the form factors.

Our results for the electric, $G_E$, and magnetic, $G_M$, form factors
for the first even-parity excitation of the proton, $p'$, are
displayed in Fig.~2 where we have easy to identify plateaus in both
cases.  Noting that the terms contributing to this result are usually
hidden by the dominant ground state within a standard unprojected
correlator, the ability to extract a result and for the duration which
we observe highlights the enormous power that the variational method
offers.

\begin{figure}[t]
\centering
\hspace*{-5pt}
\includegraphics[width=0.49\textwidth]{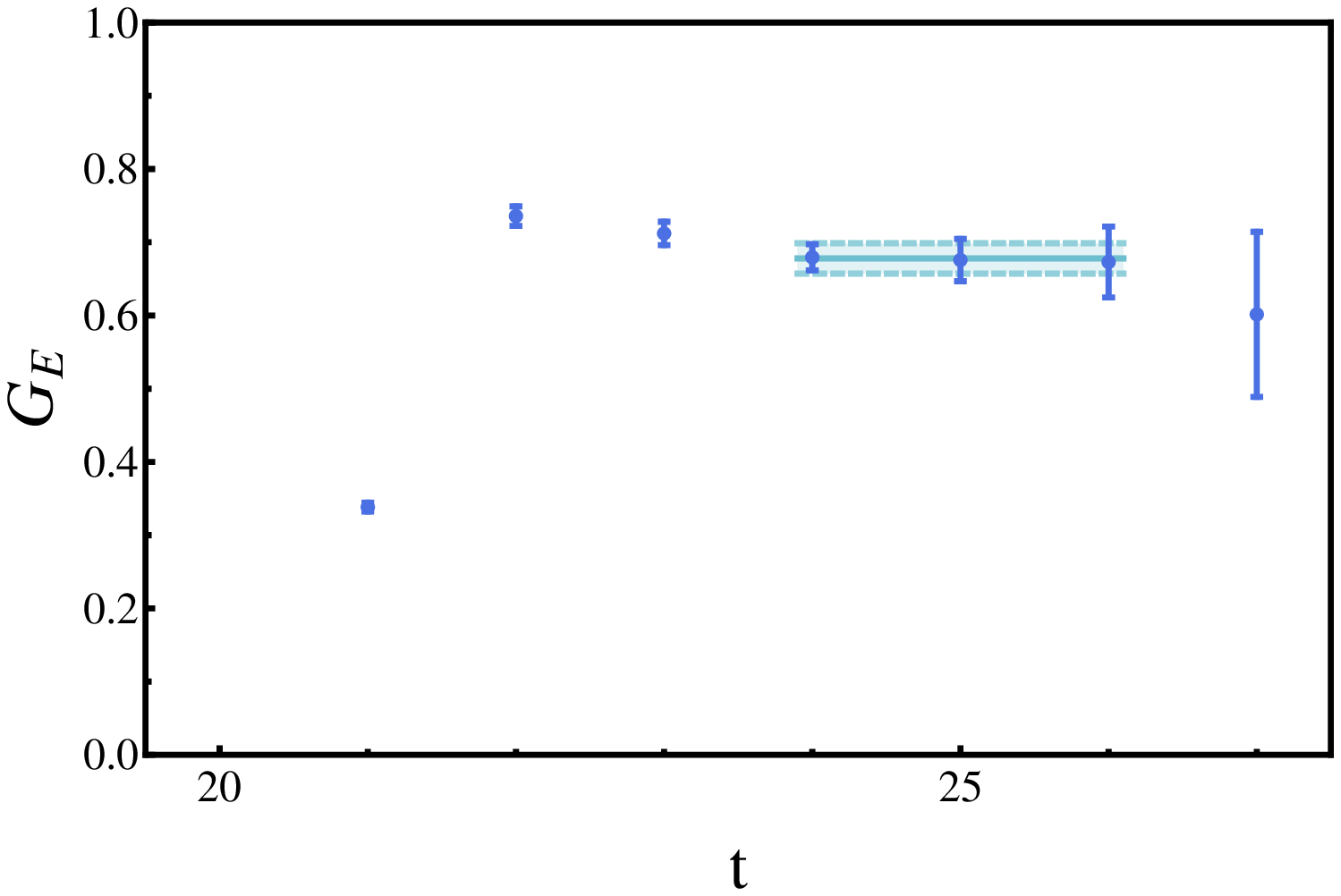}
\includegraphics[width=0.49\textwidth]{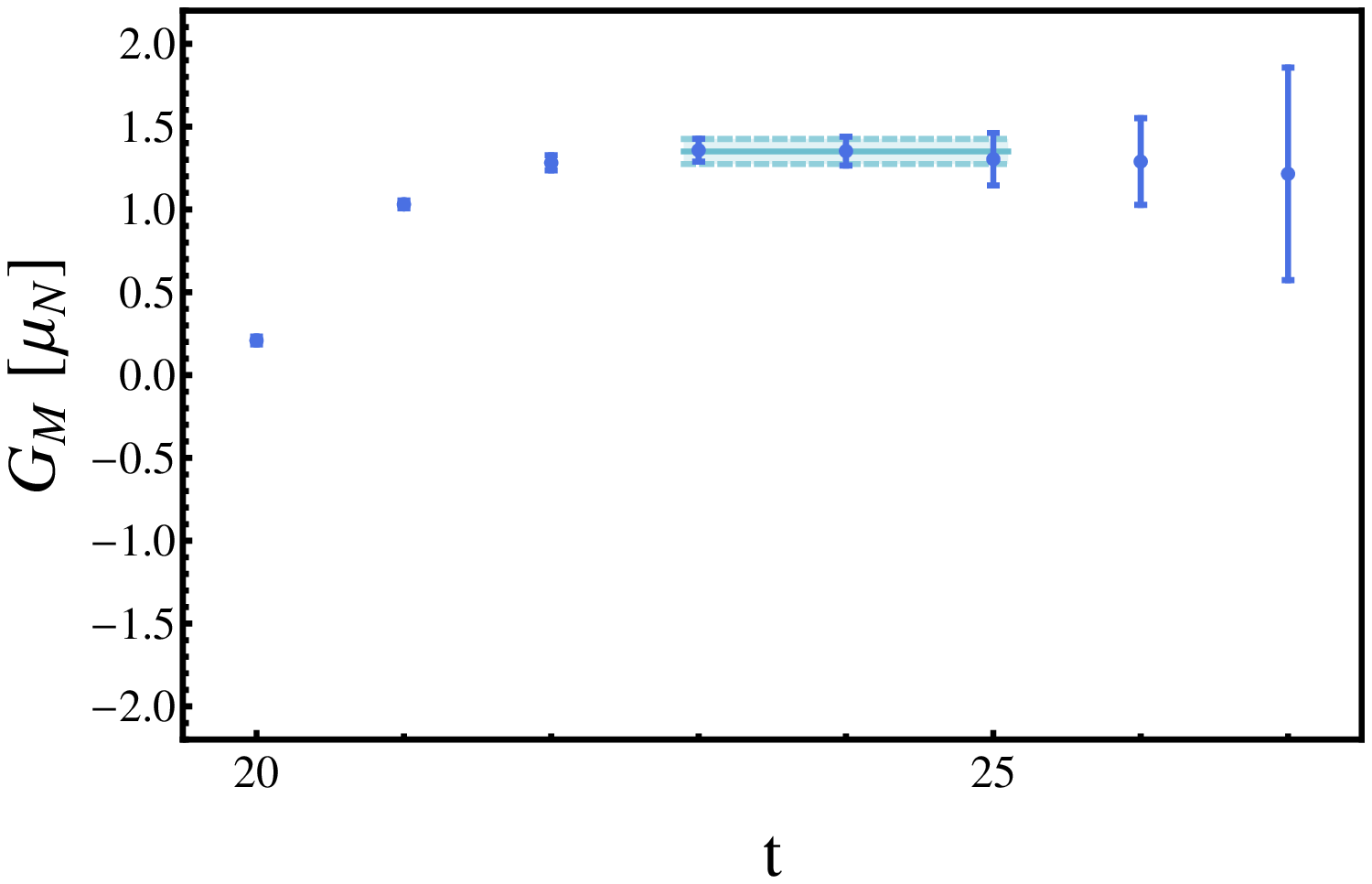}
\caption{Ratios of two- and three-point correlation functions
  providing the electromagnetic form factors of the first even-parity
  excitation of the proton at $Q^2 = 0.146$ GeV${}^2$.  The electric
  form factor, $G_E$, (left) and the magnetic form factor, $G_M$,
  (right) are illustrated.  Plateaus commence as early as two time
  slices following the current insertion at $t_S = 21$.  }
\end{figure}

Fig.~3 illustrates our results for the squared charge radii and
magnetic moments of the proton, $\Delta^+$ and the first even-parity
excitation of the proton, $p'$.  There are three key observations that we
will highlight in the following.

Firstly, the proton and $\Delta^+$ baryon have very similar form
factors across the four masses considered.  In
Refs.~\cite{Cloet:2003jm} for example, it was shown that this
behaviour is expected for the magnetic moment where cancellations of
the $\Delta^{++}-\pi^-$ and $n-\pi^+$ loop corrections to the
$\Delta^+$ leave a similar chiral behaviour for the proton and
$\Delta^+$.  Our result invites further investigation near the
physical point where the opening of the $N \pi$ threshold will change
the interplay between these contributions.

The second key observation is that the first even-parity excitation of
the proton is significantly larger than the ground state.  This is
consistent with the identification of this state being a radial
excitation of the proton through an examination of the three-quark
wave function of this state \cite{Roberts:2013}.  However, it should
also be noted that at the two values of $m_{\pi}^2$ considered herein,
the energies of these first even-parity excitations of the nucleon sit
close to the $N \pi$ $P$-wave scattering threshold.  One cannot rule
out that we may be probing an eigenstate dominated by a two-particle
$N \pi$ scattering component.  As the $N \pi$ interaction is
attractive and our box volume is 2.9 fm, the RMS-diameter of $\sim
1.7$ fm seen at the lighter of the two ensembles considered may be
reflecting an underlying multi-particle structure.

Perhaps the most interesting observation is that the magnetic moment
of the first even-parity excitation, $p'$, is similar to that of the
proton and $\Delta^+$.  While the most naive quark models would draw
on the $2S$ nature of the state and predict a magnetic moment equal to
the ground state magnetic moment of the proton, one might have
anticipated interesting effects associated with $P$-wave
multi-particle orbital angular momentum contributions to this state.

One might also expect effects associated with the increased mass of
the excited state.  Referring to Eq.~\eqref{mm}, we note that the
magnetic moment is suppressed by the hadron mass which appears in the
natural magneton of the moment.  However this effect has been
compensated by changes in the $Q^2$ dependence of the form factor
between the ground and excited state reflected in the electric form
factors.

Thus, the predictions of the naive constituent quark model in the
charge-symmetric limit,
$
\mu_p = \mu_{\Delta^+} = \mu_{p'} \, ,
$
have emerged from the complexities of quantum field theory.  As
predicted in the AccessQM model of Ref.~\cite{Cloet:2002eg} one will
have to wait for simulations in the light quark-mass regime of QCD to
see its rich structure.

\begin{figure}[t]
\centering
\hspace*{-5pt}
\includegraphics[width=0.49\textwidth]{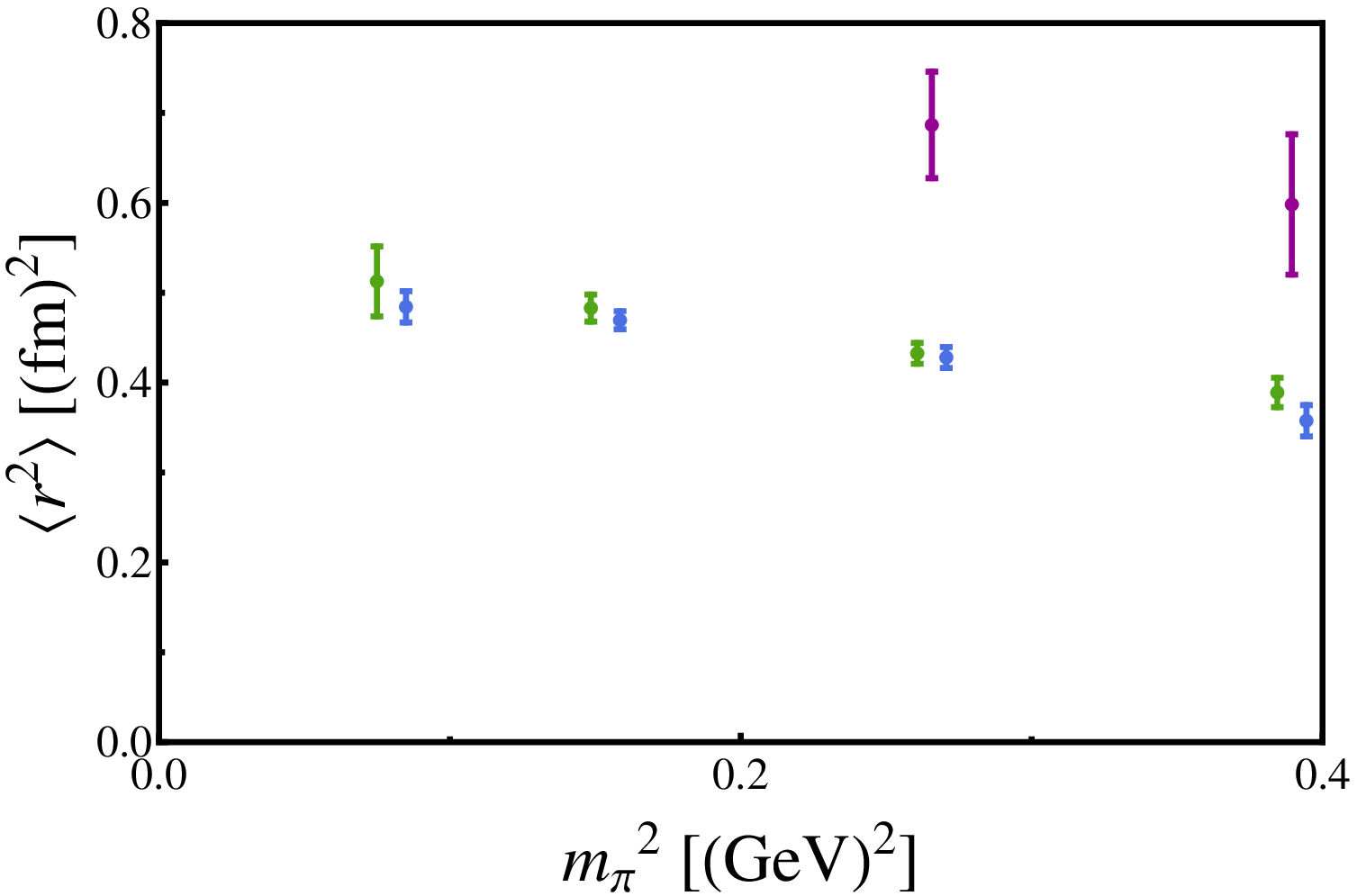}
\includegraphics[width=0.49\textwidth]{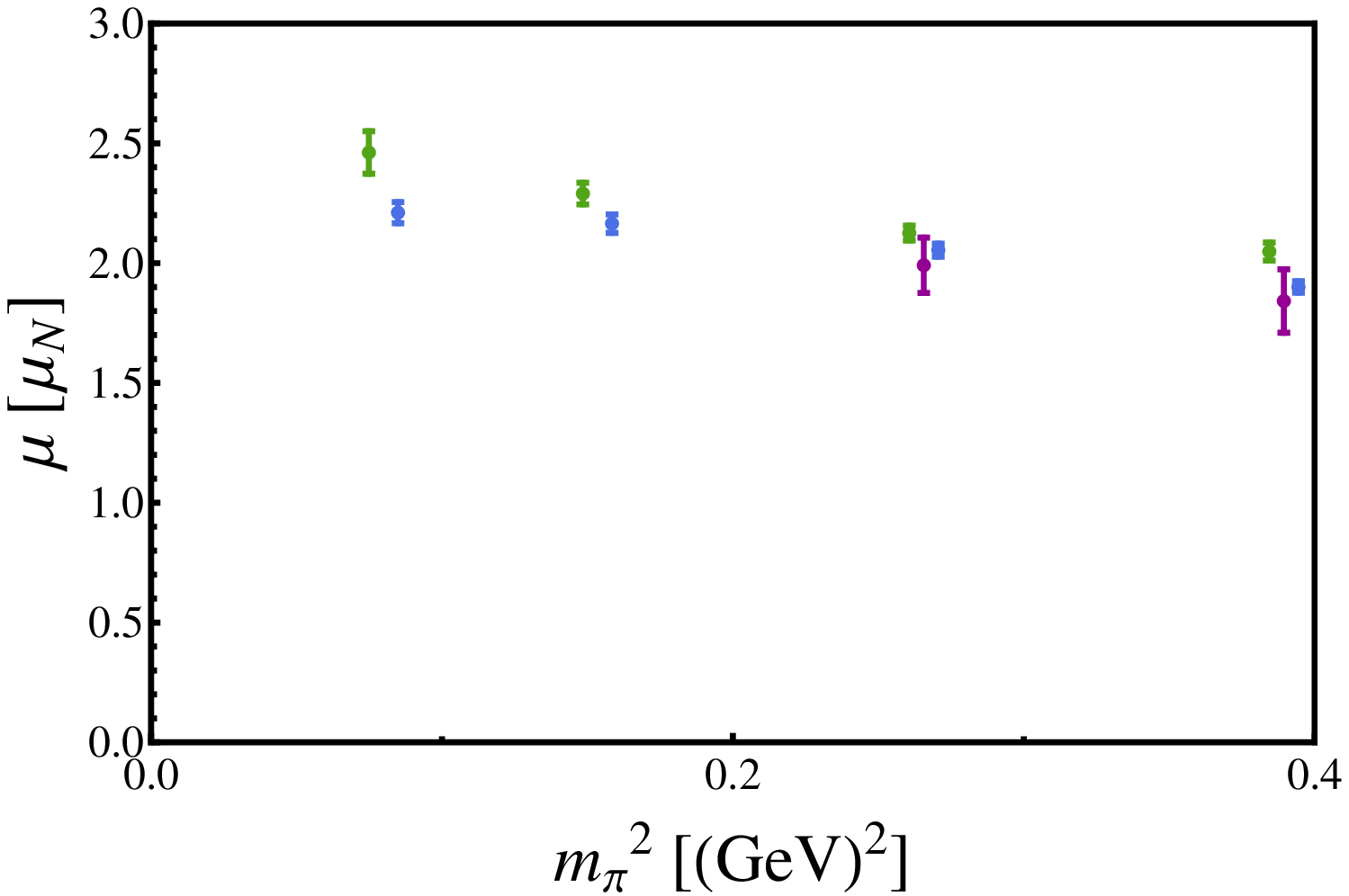}
\caption{Left: The squared charge radius, $\langle r^2_E \rangle$, for
  the proton (blue), $\Delta^+$ (green) and the first even-parity
  excitation of the proton, $p'$ (purple) in units of (fm)${}^2$.
  Right: The magnetic moment, $\mu$, for the proton (blue), $\Delta^+$
  (green) and the first even-parity excitation of the proton, $p'$
  (purple) in units of $\mu_N$.  The blue and green data points
  have been offset symmetrically
  for clarity.} 
\end{figure}

\section{Conclusion}

In this work we have demonstrated the utility of the variational
method in calculating the electromagnetic form factors of excited
states of the nucleon.  It enables a robust isolation of the first
even-parity excitation of the nucleon over a wide range of Euclidean
times and makes it possible to extract the electromagnetic form
factors.  To the best of our knowledge, this is the first
determination of the electromagnetic form factors of the first
even-parity excited state of the nucleon.

Comparison with the ground state proton and $\Delta^+$ baryons
highlights the extended nature of this excitation.  At the same time,
the predictions of early simple constituent quarks models for baryon
magnetic moments are seen to emerge from the complex interactions of
QCD.  Furthermore, we observed striking similarities between the
proton and $\Delta^+$ properties as predicted in the AccessQM model
\cite{Cloet:2002eg}.  

Extending this work to lighter masses near the $\Delta$ decay
threshold holds the promise of exposing interesting dynamics.  It will
also be interesting to explore the electromagnetic structure of
odd-parity excitations of the nucleon \cite{Mahbub:2012ri}.



\begin{thebibliography}{99}


\bibitem{Owen:2012:PLB}
  B.~J.~Owen, J.~Dragos, W.~Kamleh, D.~B.~Leinweber, M.~S.~Mahbub, B.~J.~Menadue and J.~M.~Zanotti,
  \emph{Variational Approach to the Calculation of gA},
  Phys.\ Lett.\ B {\bf 723}, 217 (2013)
  [arXiv:1212.4668 [hep-lat]].

\bibitem{Owen:2012:POS}
  B.~Owen, W.~Kamleh, D.~B.~Leinweber, S.~Mahbub and B.~Menadue,
  \emph{Correlation matrix methods for excited meson form factors in full QCD},
  PoS LATTICE {\bf 2012}, 173 (2012).

\bibitem{Leinweber:2004} 
  D.~B.~Leinweber, W.~Melnitchouk, D.~G.~Richards, A.~G.~Williams and
  J.~M.~Zanotti,
  \emph{Baryon spectroscopy in lattice QCD},
  Lect.\ Notes Phys.\  {\bf 663}, 71 (2005)
  [nucl-th/0406032].

\bibitem{Hedditch:2007} 
  J.~N.~Hedditch, W.~Kamleh, B.~G.~Lasscock, D.~B.~Leinweber, A.~G.~Williams and J.~M.~Zanotti,
	\emph{Pseudoscalar and vector meson form-factors from lattice QCD},
  Phys.\ Rev.\ D {\bf 75}, 094504 (2007)
  [hep-lat/0703014 [HEP-LAT]].

\bibitem{Leinweber:1990} 
  D.~B.~Leinweber, R.~M.~Woloshyn and T.~Draper,
  \emph{Electromagnetic structure of octet baryons},
  Phys.\ Rev.\ D {\bf 43}, 1659 (1991).

\bibitem{Leinweber:1992} 
  D.~B.~Leinweber, T.~Draper and R.~M.~Woloshyn,
  \emph{Decuplet baryon structure from lattice QCD},
  Phys.\ Rev.\ D {\bf 46}, 3067 (1992)
  [hep-lat/9208025].

\bibitem{Boinepalli:2006xd} 
  S.~Boinepalli, D.~B.~Leinweber, A.~G.~Williams, J.~M.~Zanotti and
  J.~B.~Zhang,
  \emph{Precision electromagnetic structure of octet baryons in the
  chiral regime},
  Phys.\ Rev.\ D {\bf 74}, 093005 (2006)
  [hep-lat/0604022].

\bibitem{Mahbub:2009} 
  M.~S.~Mahbub, A.~O.Cais, W.~Kamleh, B.~G.~Lasscock, D.~B.~Leinweber and A.~G.~Williams,
  \emph{Isolating Excited States of the Nucleon in Lattice QCD},
  Phys.\ Rev.\ D {\bf 80}, 054507 (2009)
  [{\tt arXiv:0905.3616 [hep-lat]}].

\bibitem{Mahbub:2010} 
  M.~S.~Mahbub {\it et al.}  [CSSM Lattice Collaboration],
	\emph{Roper Resonance in 2+1 Flavor QCD},
  Phys.\ Lett.\ B {\bf 707}, 389 (2012)
  [arXiv:1011.5724 [hep-lat]].

\bibitem{Aoki:2008} 
  S.~Aoki {\it et al.}  [PACS-CS Collaboration],
  \emph{2+1 Flavor Lattice QCD toward the Physical Point},
  Phys.\ Rev.\ D {\bf 79}, 034503 (2009)
  [{\tt arXiv:0807.1661 [hep-lat]}].

\bibitem{Beckett:2011} 
  M.~G.~Beckett, B.~Joo, C.~M.~Maynard, D.~Pleiter, O.~Tatebe and T.~Yoshie,
  \emph{Building the International Lattice Data Grid},
  Comput.\ Phys.\ Commun.\  {\bf 182}, 1208 (2011)
  [{\tt arXiv:0910.1692 [hep-lat]}].

\bibitem{Cloet:2003jm} 
  I.~C.~Cloet, D.~B.~Leinweber and A.~W.~Thomas,
	\emph{Delta baryon magnetic moments from lattice QCD},
  Phys.\ Lett.\ B {\bf 563}, 157 (2003)
  [hep-lat/0302008].

\bibitem{Roberts:2013} 
  D.~S.~Roberts, W.~Kamleh and D.~B.~Leinweber,
  \emph{Wave Function of the Roper from Lattice QCD},
  arXiv:1304.0325 [hep-lat].

\bibitem{Cloet:2002eg} 
  I.~C.~Cloet, D.~B.~Leinweber and A.~W.~Thomas,
  \emph{Simple quark model with chiral phenomenology},
  Phys.\ Rev.\ C {\bf 65}, 062201 (2002)
  [hep-ph/0203023].

\bibitem{Mahbub:2012ri} 
  M.~S.~Mahbub {\it et al.},
  \emph{Low-lying Odd-parity States of the Nucleon in Lattice QCD},
  Phys.\ Rev.\ D {\bf 87}, 011501 (2013)
  [arXiv:1209.0240 [hep-lat]].

\end{thebibliography}
\end{document}